# Enhanced ferromagnetism in artificially stretched lattice in quasi two-dimensional $Cr_2Ge_2Te_6$

Hiroshi Idzuchi[1-4]†*, Andres E Llacsahuanga Allcca[3]†, Anh Khoa Augustin Lu[5,6], Mitsuhiro Saito[1,7], Michel Houssa[8], Ruishen Meng[8], Kazutoshi Inoue[1], Xing-Chen. Pan[1], Katsumi Tanigaki[1,9], Yuichi Ikuhara[1,7], Takeshi Nakanishi[5], and Yong P Chen[1-3,10,11]*

[1] *WPI Advanced Institute for Materials Research (AIMR), Tohoku University, Sendai 980-8577, Japan.*
[2] *Center for Science and Innovation in Spintronics (CSIS), Tohoku University, Sendai 980-8577, Japan.*
[3] *Department of Physics and Astronomy, Purdue University, West Lafayette, Indiana 47907, USA.*
[4] *Department of Physics, The University of Tokyo, Bunkyo-ku, Tokyo 113-0033, Japan.*
[5] *Mathematics for Advances Materials Open Innovation Laboratory, National Institute of Advanced Industrial Science and Technology, Sendai 980–8577, Japan.*
[6] *International Center for Materials Nanoarchitectonics, National Institute for Materials Science, Tsukuba 305-8568, Japan*
[7] *Institute of Engineering Innovation, The University of Tokyo, Tokyo 113-8656, Japan.*
[8] *Department of Physics and Astronomy, Katholieke Universiteit Leuven, Leuven 3001, Belgium*
[9] *Beijing Academy of Quantum Information and Sciences, Beijing 100193, China.*
[10] *School of Electrical and Computer Engineering and Birck Nanotechnology Center and Purdue Quantum Science and Engineering Institute, Purdue University, West Lafayette, Indiana 47907, USA.*
[11] *Institute of Physics and Astronomy and Villum Centers for Dirac Materials and for Hybrid Quantum Materials and Devices, Aarhus University, 8000 Aarhus-C, Denmark.*

**Abstract**
In the fundamental understanding of magnetic interactions between atoms in solids, the crystal lattice is one of the key parameters. As the effective tool for controlling the lattice using tensile stress is limited, there are only few demonstrations of the control in magnetic properties with expanding the lattice structure. Here, we observe that the Curie temperature ($T_c$) of quasi two-dimensional $Cr_2Ge_2Te_6$ with NiO overlayer doubles from ~60 K to ~120 K, describe a clear correlation of magnetic properties with lattice expansion, which is characterized by several probes and computational approaches, and address on the mechanisms leading to the increase in $T_c$ via the change in exchange interactions.

-----------------------------------------------------------------------------------------------------------------

† These two authors contributed equally *Corresponding author. Email: hiroshi.idzuchi@phys.u-tokyo.ac.jp, chen276@purdue.edu.

Magnetic interactions at the atomic level play a central role in magnetism. The recent rise of two-dimensional van der Waals (vdW) magnetic materials offers the possibility to study the magnetic interactions thanks to their high crystallinity sample and the possibility to study various thicknesses [1,2], whose characteristics can be easily accessed by a number of probes with spatial resolution such as Raman spectroscopy [3-5]. One of the most important indications of magnetic interactions is the Curie temperature ($T_C$). Together with the practical motivation to increase $T_C$, the relation between magnetic interactions and $T_C$ has been widely studied in vdW magnets. For example, the change in magnetic interactions with the electric structure and carrier concentration was studied by electric gating (in particular in the structure of a field-effect transistor), which changes the hysteresis curve for a localized magnetic system of $Cr_2Ge_2Te_6$ without any significant change in $T_C$, while the $T_C$ increases from 205 K to above room temperature in the case of an itinerant magnetic system of $Fe_3GeTe_2$ in similar structure [6,7]. Historically, external pressure has been used for tailoring the lattice parameter of magnetic systems. This approach was applied to both iterant and localized magnetic systems, but it was found that $T_C$ of prototypical ferromagnet $CrI_3$ only slightly increased from 44 K to 48 K [8]. While an external pressure could be used for shrinkage, it cannot be used for an expansion of the lattice and the studies in small size samples are rather limited. In the present study, we present a $Cr_2Ge_2Te_6$/NiO heterostructure, forming wrinkled structure to induce in-plane tensile strain, that can achieve nearly twice the $T_C$ of bulk and as-cleaved $Cr_2Ge_2Te_6$. Such hetero-interface provides a new tuning knob for controlling the order parameters to modify the electronic structure, which cannot be accessed by conventionally used pressure-cell.

   When an in-plane tensile force is applied to one part of a 2D layer (Fig.1a) while other part keeps a constant length, a wrinkle forms in the stretched part in order to conserve the total number of atoms (Fig.1b). From such deformation introducing a gap, one would expect that the optical contrast of 2D material flakes on a Si/SiO$_2$ substrate to significantly changes across the wrinkle, given that the changing optical path which is introduced by the gap modifies optical interference condition [9]. We prepared a sample from a single crystal of $Cr_2Ge_2Te_6$, which was formed on a silicon substrate by mechanical exfoliation. A NiO layer was formed onto $Cr_2Ge_2Te_6$ by sputtering. Details of the sample preparation are described in the previous report [10,11]. Figures 1c and 1e show the optical micrograph of $Cr_2Ge_2Te_6$ flake on a Si/SiO$_2$ substrate (here the thickness of SiO$_2$ was 285 nm) with and without a NiO overlayer (cross-section is shown in Fig,1d), respectively (note NiO itself is semitransparent). $Cr_2Ge_2Te_6$ with NiO flake shows differentiated contrast at some positions, which are correlated with the high height positions as seen by atomic force microscope topography (Fig.1e-1g) consistent with a wrinkle structure.

   The wrinkle structure is further examined by the cross-sectional view of the sample. Figure 1h shows a scanning transmission electron microscopy (STEM) image of $Cr_2Ge_2Te_6$/NiO on a silicon substrate supported by a focused ion beam (FIB) grid [12]. This shows that $Cr_2Ge_2Te_6$/NiO is indeed detached from the Si/SiO$_2$ layer and a wrinkle structure is clearly visible. The area between the wrinkled layer and SiO$_2$ (the white area near the center of Fig.1h) was examined by energy dispersive X-ray spectroscopy (STEM-EDS), but no noticeable spectrum from elements was detected, indicating that no material exist in this area and $Cr_2Ge_2Te_6$/NiO was separated by Si/SiO$_2$ layer. From the consistent experimental observations by optical microscopy, atomic force microscopy, and scanning transmission electron microscopy of the cross-section of our samples, we concluded that wrinkles were formed in the $Cr_2Ge_2Te_6$/NiO heterostructure. The NiO thickness ranged from 20 to 100 nm. The heterostructures were studied for various $Cr_2Ge_2Te_6$ thickness values, up to bulk limit. The formation of wrinkle was observed for all $Cr_2Ge_2Te_6$ thickness. For thick $Cr_2Ge_2Te_6$ samples, the portion close to the interface NiO/ $Cr_2Ge_2Te_6$ deforms and delaminates from the rest of the sample. Graphite, graphene, and $MoS_2$ [13,14] are just a few examples of 2D vdW materials in which the wrinkle structure has so far been noted. In contrast to typical situation where a flake forms on a flexible substrate and the substrate is deformed [15], our case involved an overlayer (NiO) that induces a change in lattice spacing.

The magnetic properties of the $Cr_2Ge_2Te_6$/NiO heterostructure were characterized by means of magneto-optical Kerr effect (MOKE) [12]. Figure 1i shows the variation in $T_C$ for three typical cases: two types of $Cr_2Ge_2Te_6$/NiO (flakes with and without wrinkle) and $Cr_2Ge_2Te_6$ without NiO layer. A flake of $Cr_2Ge_2Te_6$/NiO with wrinkle structure exhibits a higher $T_C$ for various thicknesses values, with up to nearly twice of $Cr_2Ge_2Te_6$ in the absence of the NiO layer. The increase in $T_C$ for $Cr_2Ge_2Te_6$/NiO has been reported in a previous work, but the underlaying mechanism remained inconclusive as the influence of the wrinkle structure was unknown at that time [10].

As the wrinkle structure breaks the translational symmetry, the spatial distribution of Raman and MOKE characteristics are examined. Figure 2b shows the Raman spectrum at four different positions of a representative flake (optical micrograph and the positions are shown in Fig.2a). Five characteristic peaks are found in the 80-310 cm$^{-1}$ range at room temperature [4]. The position of the peaks in the spectrum changes with the position in the flake, even for areas of the same thickness, indicating that the strain varies with the position of the flake. The peak near 235 cm$^{-1}$ (110 cm$^{-1}$) is ascribed to an (two) in-plane vibration mode(s), whereas the other three peaks are assigned to other types of vibrations, i.e., an out-of-plane mode and a combination of the two modes. The MOKE measurement at the same position (Fig. 2c) indicates that $T_C$ also varies with the position. A correlation of Raman peak and $T_C$ is illustrated in Fig.2d, wherein the Raman-peak-position map together with $T_C$ indicate that when the Raman shift is lower in frequency (corresponding to tensile strain [16]), $T_C$ becomes higher.

Next, we address the general trend over sample-to-sample variation in the correlation of magnetism and stretched lattice. We noticed that the wrinkle formation does not always happen in every exfoliated $Cr_2Ge_2Te_6$ flake; however, once it does, it appears for all ranges of thicknesses up to the bulk crystal limit [17]. This could be contributed by the rather weak interlayer coupling inside the crystal. The wrinkle formation was also occasionally observed in flakes in the middle of the cooling process in the cryostat, where strain in such a stochastic behavior was further examined by studying 57 positions on 27 flakes with and without NiO [12]. The Raman frequency is distinctively smaller for the flakes with wrinkle than those for without wrinkle regardless of the presence of NiO layer (Figure 3a). Assuming a single strain value in each spot, the shift of two Raman mode should correlate, which is clearly exemplified in Fig.3a for $E_{g1}$ (~111-112 cm$^{-1}$) and $E_{g4}$ (~ 234-235 cm$^{-1}$) peaks (please refer to SI for a combination with $E_{g2}$ (~137 cm$^{-1}$) mode). The correlation between $T$c and Raman peaks is further examined on four flakes that exhibit both in-plane Raman modes. Taking into account that a 1% lattice shrinkage reportedly corresponds to a shift about 0.3 – 0.5 cm$^{-1}$ for the peak around 110 cm$^{-1}$ (values extracted from pressure cell experiment [18]), our data indicates a ratio of 5 - 8 K (8 – 12%) change of $T$c with a lattice expansion of 1 % ($\approx$16 K/cm$^{-1}$). The sign change in wave number and in lattice parameter is consistent with the results obtained from first-principles studies on monolayer $Cr_2Ge_2Te_6$ [19].

Magnetism in $Cr_2Ge_2Te_6$ mainly relies on the magnetic interactions between Cr atoms, which in a quasi-2D triangular lattice system has been studied in $M$CrS$_2$ systems, where $M$ is a monovalent metal among Li, Na, K, Ag and Au [20,21]. Specifically, three unpaired electrons from Cr form triplet $t_{2g}$ orbitals with a lower energy than doublet $e_g$ states. The direct overlap between different $t_{2g}$ orbitals of neighboring Cr ions, which are along Cr-Cr direction, give rise to strong anti-ferromagnetic (AFM) exchange interactions, while strongly decreasing with an increasing Cr-Cr distance. This competes with ferromagnetic (FM) interactions that are caused by virtual hopping from the occupied $t_{2g}$ shell of one Cr to the empty $e_g$ shells of another Cr. This explains why the AFM (FM) nature is more pronounced in small (large) $M$ in $M$CrS$_2$ compounds. In $Cr_2Ge_2Te_6$, we expect that the increase in Cr-Cr distance reduces the AFM exchange interaction, resulting in the net increase of the FM strength. The increased FM interaction in the stretched lattice is further confirmed by density functional theory (DFT) calculation and Monte Carlo simulations, as shown in Fig.3c. As tensile strain increases, the exchange energy thus $T_C$ and the magneto anisotropy increase [22]. Although similar calculations were previously reported in the atomically thin limit [19,23], we computed the increase of $T_C$ for the case of a thick flake, which corresponds to configuration of our experiment. Notably, for small strain values, a tensile strain of 1% leads to an increase in $T_C$ of $\approx$17 % in reasonable agreement with the experimental data.

The strain inferred by the Raman spectrum and magnetic properties, was further characterized by cross-sections of the $Cr_2Ge_2Te_6$/NiO ($t_{NiO}$ = 50 nm) heterostructure using STEM [24]. Figures 4a and 4b display periodic black-and-white contrast views of the bottom region, which is caused by the electrons scattering from the $Cr_2Ge_2Te_6$ layers in the direction along the crystal axis perpendicular to the zigzag direction of Cr-Ge-Cr. A representative trace of the contrast gives a period of $\approx$ 0.19 nm (Figs.4b and 4c [25]), in a good agreement with previous reported values for a $Cr_2Ge_2Te_6$ crystal (0.197 nm [20]). Interestingly, the $Cr_2Ge_2Te_6$ layers located near the $Cr_2Ge_2Te_6$/NiO interface (L$_2$ region) have a larger period (0.194 nm) than those located deeper (0.189 nm). Such a large period is also visible in the Fourier transformed image [25] after summing up the contrast of each layer (Fig.4c), indicating a characteristic thickness of ≈12 layers (8 nm, not including the regions that do not show the periodic contrast ($L_1$ region)). Within the thickness of $Cr_2Ge_2Te_6$ on $SiO_2$/Si ≈ 27 nm in this specimen, apart from the first 12 layers ($L_1$ and $L_2$ regions), the rest of the layers (Reference region) towards $SiO_2$ showed no notable change in the in-plane lattice parameters (note we concentrated on samples with relatively thin flake for Raman and MOKE measurements in figs. 2 and 3 in order for the signal sensitive to the strained layer). An energy dispersive X-ray spectroscopy image of this specimen (Fig.4d) shows that there is no notable doping in the $L_2$ region. However, at the same time, we also notice that a small amount of Ni atoms (~1%) are substituted up to ~10 nm in depth. First-principles simulations of $Cr_2Ge_{2-1/8}Te_6$-$Ni_{1/8}$ shows that wiggle or buckling appears as a consequence of structure relaxations. This is consistent with our characterization of the specimen [26].

In conclusion, we demonstrated an artificially stretched lattice in a crystalline $Cr_2Ge_2Te_6$ sample, by characterizing magnetic properties, crystal structure and electronic structure by employing a combination of experimental and theoretical methods. Our present results open the possibility of engineering the lattice in 2D vdW materials. While lattice engineering in magnetic materials has been thus far rather limited, our new approach by means of hetero-interface structure of 2D vdW material, of which interface we do not limit within 2D materials but include non 2D materials, provides an effective tool for strain engineering to control the electronic structure over a wide temperature range [27]. Our methodology can further be extended to engineer the lateral compressive or tensile strain on any 2D layer, which may apply to forming a quantum emitter [28], confining the potential or exhibiting moiré effect, as well as developing further functionalities.

**Acknowledgements**

This work was supported in part by AIMR and its "fusion" research program, under World Premier International Research Center Initiative (WPI), the Ministry of Education, Culture, Sports, Science and Technology (MEXT), Japan, and by Grant-in-Aid for Scientific Research, JSPS KAKENHI (Grant Numbers, 20K14399 and 22H01896), and JST-CREST (JPMJCR18T1).

**Figures**

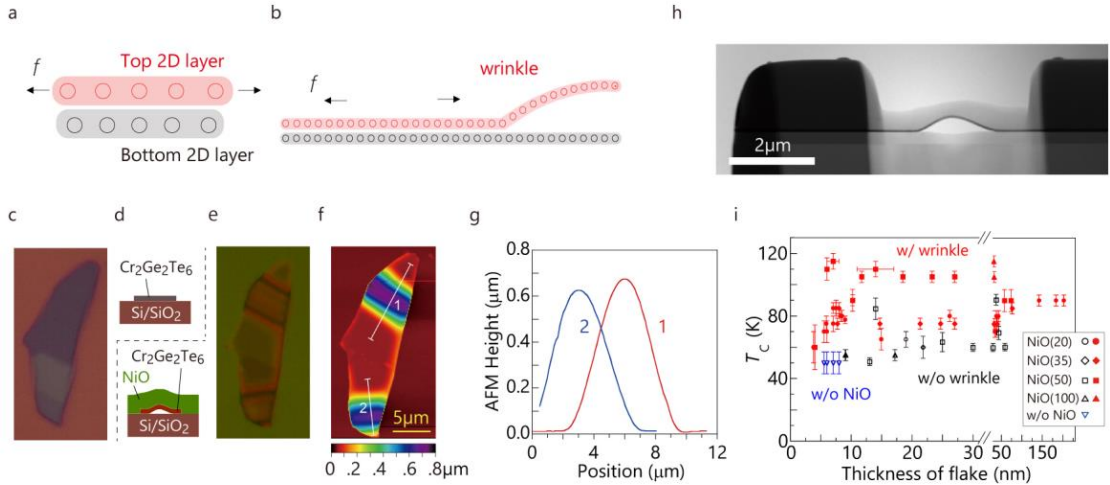

**FIG. 1. Wrinkle formation in Cr$_2$Ge$_2$Te$_6$/NiO a**, Mechanism to form wrinkle in 2D layers (red and black filled regions). Each circle represents an atom. In the top layer, an in-plane tensile force *f* is applied and result in a laterally displaced from the original position (indicated by the bottom layer in black). **b**, In other regions, the low density may induce a deformation of the layer (such as wrinkles), especially when the interlayer coupling between the sheets is weak, which is the case in 2D materials. **c-e**, Optical micrograph of (**c**) the Cr$_2$Ge$_2$Te$_6$ flake and (**e**) the Cr$_2$Ge$_2$Te$_6$ flake with NiO layer placed on a silicon substrate. The cross-sections of (**c**) and (**e**) are schematically shown at the top and bottom of (**d**), respectively. **f**, Atomic force micrograph (AFM) of the sample shown in (**e**). **g**, Height of the flake along the lines indicated in (**f**), measured by AFM. **h**, STEM image of Cr$_2$Ge$_2$Te$_6$/NiO on a silicon substrate, supported by a FIB grid (which is appears as the dark gray blocks on the left and right). **i**, Curie temperature of Cr$_2$Ge$_2$Te$_6$ flake with (red and black) and without (blue) NiO layer. The results for flakes with (resp. without) wrinkles are indicated by closed red (resp. open black) symbols. The symbols reflect the NiO thickness: 20 nm (circle), 35 nm (diamond), 50 nm (rectangle), and 100 nm (triangle).

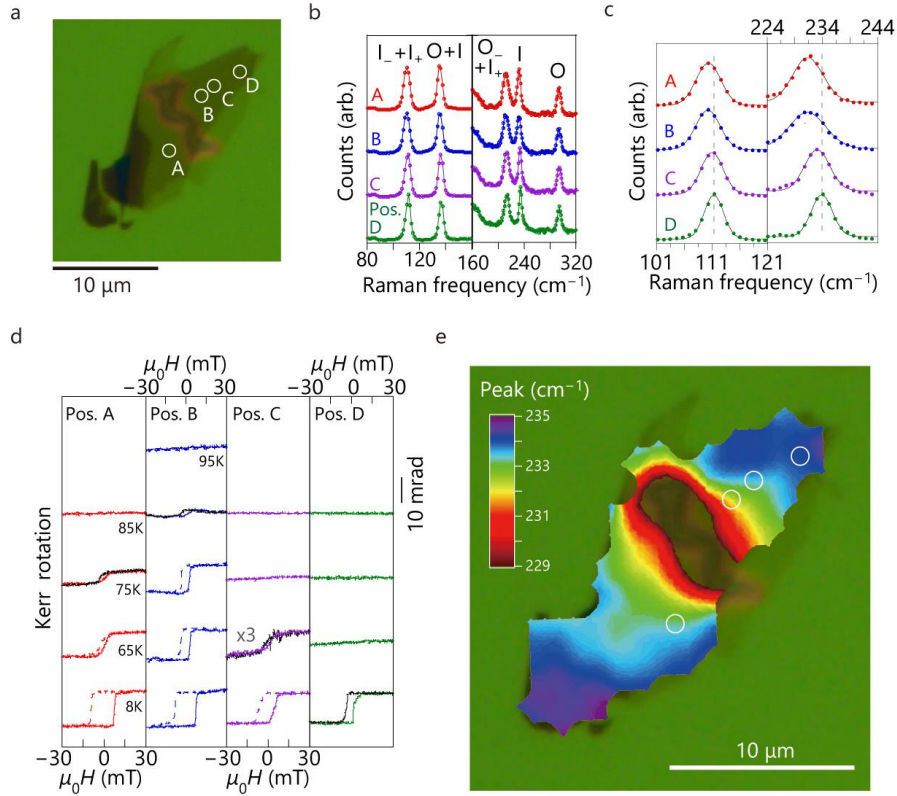

**FIG. 2. Raman spectrum and magnetic properties of a wrinkled flake. a,** Optical micrograph of a $Cr_2Ge_2Te_6$ flake with a 50-nm-thick NiO overlayer. Wrinkles appear as wavy patterns near the center of the flake. Circles (labeled as A, B, C and D) represent the positions. **b,c,** Raman spectrum of the flake at positions A (red), B (blue), C (purple), and D (green), in the 80 – 320 cm$^{-1}$ range. The labels "I"("O") represent in-plane (out-of-plane) mode. For the nearly degenerated peaks, the subscript + (−) represents a mode with a slightly higher (lower) frequency with respect to the center of the peaks. According to theoretical calculations, five peaks are expected in this frequency range (I_+I_+, O+I, O_-+I_+, I, and O), with three of which are nearly doubly degenerated modes. The solid line connecting the data points in (**b**) represents fitted curves. In (**c**), the dotted lines indicate the Raman peak for position D in the 101 – 121 cm$^{-1}$ and 224 – 244 cm$^{-1}$ ranges. **d,** Kerr rotation at positions A (red), B (blue), C (purple) and D (green), from left to right. The hysteresis curves for magnetic fields (in perpendicular to the sample plane) are presented for several temperatures (8 K, 65 K, 75 K, 85 K and 95 K). The curves are shifted vertically for better clarity. Sweeps from positive to negative value are indicated by dotted or black lines. The scale bar indicates 10 mrad. **e,** Spatial distribution of position of the peak Raman-peak near 234 cm$^{-1}$. The color indicates the position of the peak (Raman frequency), as indicated by the color bar. The contour plot was generated by interpolating values measured every 0.8 μm.

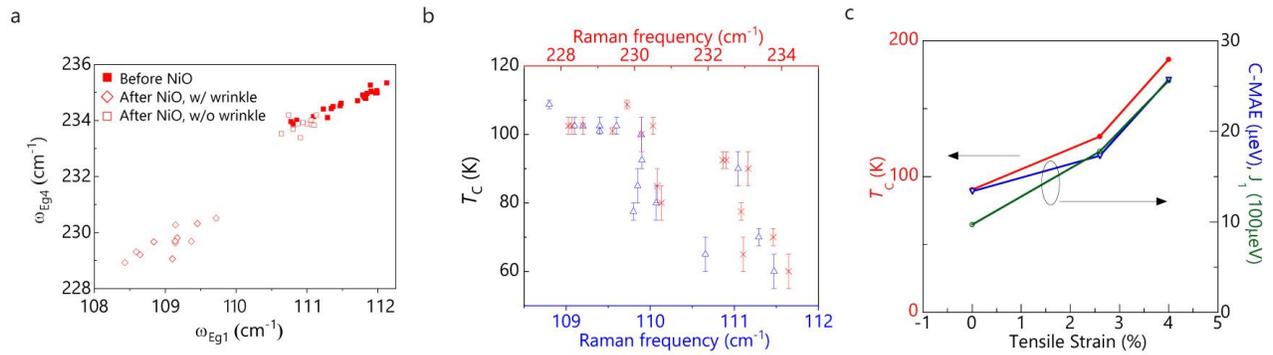

**FIG. 3. Correlation between tensile strain and enhanced magnetism. a,** The correlation of Raman modes which include only in-plane vibration. Samples (before) after forming NiO which does not show wrinkles are indicated as (open) solid rectangle and circle. Samples after forming NiO which shows wrinkle are indicated by diamond. **b,** Relationship between the Curie temperature and the in-plane Raman mode for samples with a NiO layer. **c,** Curie temperature $T_C$ (red filled circle), crystalline magneto anisotropy energy C-MAE (blue triangle), and first exchange coupling $J_1$ (green empty circle) with respect to the in-plane strain, from DFT calculations.

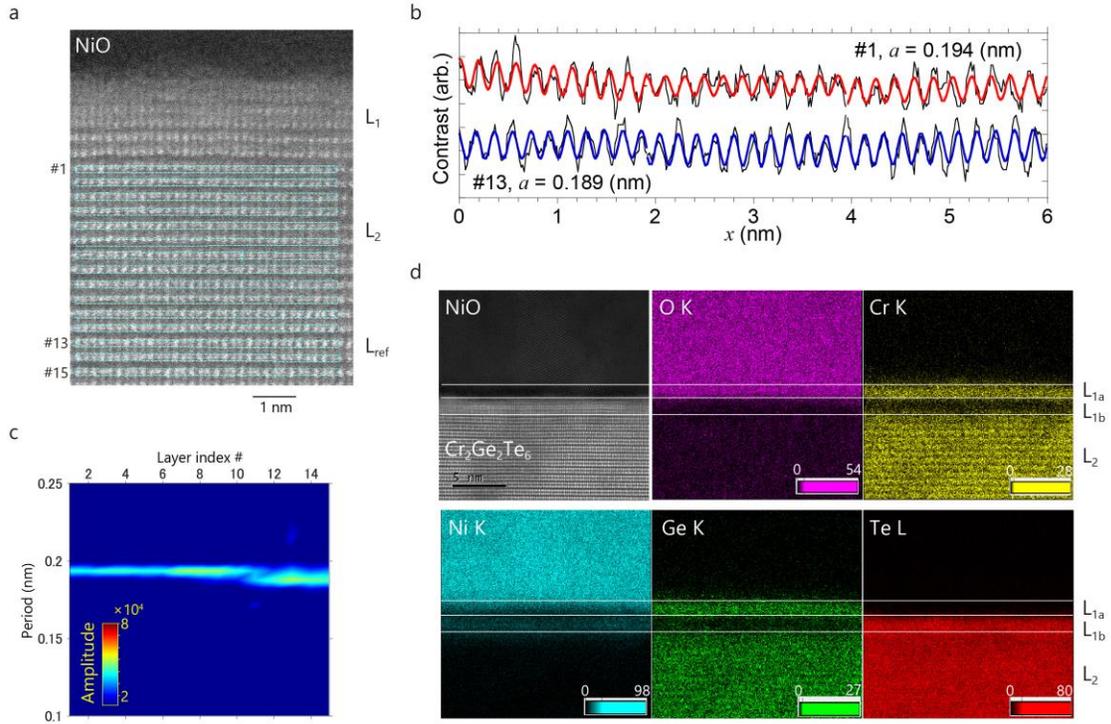

**FIG. 4. STEM image and tensile strain in Cr$_2$Ge$_2$Te$_6$ layer. a,** STEM image of Cr$_2$Ge$_2$Te$_6$/NiO. The upper dark region indicates the NiO layer. In the lower region, well-defined flat layered patterns (highlighted by blue dashed rectangles) indicate the Cr$_2$Ge$_2$Te$_6$ layers. These layers are individually numbered (#1, #2, ... #15). Between them, there is a less flat layer with less sharp contrast (labeled as L$_1$). Blue rectangles highlight the area to count the contrast in (**b**) and (**c**). L$_2$ (L$_{ref}$) is defined in (**c**). **b,** The contrast variation with lateral position in the column #1 and #13 shown in (**a**). Red and blue curves show the fitting with the sine function (periods represent 0.194 nm and 0.189 nm). **c,** Fourier transform of the highlighted region in (**a**). In (**a**), layers that are nearer (far) from #11 layer is labeled as L$_2$ (L$_{ref}$). **d,** STEM EDS mappings for O, Cr, Ni, Ge and Te atoms. First panel shows a STEM image for the area. At the interface, there is a region of intermixing, labeled as L$_{1a}$ and L$_{1b}$, the total of which corresponds to L$_1$ in (**a**).

11